\documentclass[11pt,twoside]{article} 
\usepackage{asp2004}
\usepackage{epsf}
\usepackage{psfig}
\begin{document}
\title {Spectroscopic Equilibrium of Iron in Metal-Rich Dwarfs} 

\newcommand{\teff}{T$_{\rm eff}$ }
\newcommand{\tirfm}{T$_{\rm eff}^{\rm IRFM}$}
\newcommand{\tfe}{T$_{\rm eff}^{\rm FeI}$}

\author{Jorge Mel\'endez}
\affil{Caltech, Department of Astronomy, M/C 105-24, 
1200 E. California Blvd, Pasadena, CA 91125}

\author{Iv\'an Ram\'{\i}rez}
\affil{University of Texas, Department of Astronomy,
RLM 15.306, TX 78712}

\begin{abstract}
We analyze twenty five nearby metal-rich G and late-F dwarfs
in order to verify whether the spectroscopic equilibrium (LTE)
of iron lines satisfy the observational constraints imposed 
by the Infrared Flux Method (angular diameters) and 
Hipparcos parallaxes. The atmospheric parameters
derived from iron lines (assuming LTE and employing 1D Kurucz
model atmospheres) do not satisfy simultaneously both
observational constraints, probably because classical modeling
fails to reproduce the detailed line formation of FeI lines.

\end{abstract}

\section{Introduction}

Until a few years ago, it was not possible to directly 
determine the fundamental properties of even the closest dwarf stars.  
The advent of {\it Hipparcos} has greatly improved the situation,
allowing to accurately determine the surface gravities
of nearby dwarfs. The situation is much worse for the 
effective temperatures (\teff) of unevolved stars, due to
sub-milli-arcsec level angular diameters 
(a few mas for the closest dwarfs). 
Fortunately, the stellar diameters
of about a dozen dwarfs are now available, mostly from VLTI 
interferometric measurements (Kervella et al. 2004).

Ram\'{\i}rez \& Mel\'endez (2004, 205a) have shown that the 
Infrared Flux Method (IRFM) \teff scale 
is in excellent agreement with 
measured stellar diameters of 10 FGK dwarfs in the
metallicity range -0.5 $<$ [Fe/H] $<$ +0.3 dex. However, the
temperatures obtained from spectroscopic equilibrium of iron lines
seem to be in conflict with the ones derived from the IRFM. 
The temperatures of metal-rich dwarfs 
with planets may be being overestimated by about 110 K
when the temperatures are determined from excitation
equilibrium of FeI lines (see Ram\'{\i}rez \& Mel\'endez 2004
and references therein). This is not restricted to star
with planets, since analysis of other samples results in
similar discrepancies. For example, the FeI temperatures of
thin- and thick-disk dwarf stars analyzed by Bensby et al. (2003) are
also about 110 K hotter than IRFM temperatures (Ram\'{\i}rez \&
Mel\'endez 2005a).

The source of the problem is still not clear. The culprit is
probably the neglect of NLTE effects, but errors in atomic data 
are not discarded, as well as granulation effects.
In this work we analyze a sample of nearby metal-rich
dwarfs from the S$^4$N database (Allende Prieto et al. 2004), 
which are closer than 15 pc,
therefore with extremely small errors 
in their {\it Hipparcos} parallaxes, imposing stringent constraint on 
their surface gravities ($<$ 0.01 dex), 
allowing thus to check the validity of spectroscopic gravities. 
Furthermore, the IRFM temperature scale 
is used to check the excitation temperatures determined from a 
1D LTE analysis.

\section{Analysis}
The sample stars were selected from the S$^4$N database,
and they cover the atmospheric parameters space 
5000 K $<$ \teff $<$ 6600 K, 4.0 $<$ log $g$ $<$ 4.6,
$-0.8 <$ [Fe/H] $<$ +0.5. Twenty five stars 
with relatively clean line profiles were selected.

The temperature was determined employing the IRFM
\teff scale (Ram\'{\i}rez \& Mel\'endez 2005a,b) and 
surface gravities are from {\it Hipparcos} parallaxes.
We will refer to these atmospheric parameters 
as ``physical'' parameters.

The choice of atomic data is critical,
since systematic errors in the $gf$-values could lead to
wrongly-determined atmospheric parameters. For FeI lines, 
the $gf$-values were taken from the Oxford (Blackwell et al. 1995a
and references therein) and Hannover (Bard et al. 1991,
Bard \& Kock 1994) groups. The collisional broadening constants 
were taken from Barklem et al. (2000).

It has been discussed in the literature the lack of
accurate experimental $gf$-values for FeII lines, 
and the importance of improving laboratory measurements for
accurate stellar abundance work
(e.g. Lambert et al. 1996, Gehren et al. 2001). 
The bulk of laboratory measurements are
probably correct in an absolute scale, but there are large
uncertainties in a line-by-line basis. On the other hand, theoretical
{\it relative} line intensities within a given multiplet
are highly reliable, but the absolute scale is not.
Mel\'endez \& Barbuy (2002) have taken advantage of both methods,
with relative $gf$-values obtained from theoretical calculations 
and the absolute transition probabilities for each multiplet 
were normalized with laboratory measurements.
Mel\'endez \& Barbuy (2005, in preparation) have updated their 
previous work with new laboratory and theoretical works.
This improved list of FeII lines is used in the present work.

Iron abundances were obtained from FeI and
FeII lines, employing the program MOOG (Sneden 1973)
and Kurucz model atmospheres.
First, iron abundances were determined from the 
physical parameters (IRFM + Hipparcos), 
then LTE spectroscopic equilibrium was imposed 
to obtain spectroscopic temperatures and gravities,
enforcing simultaneously
$\Delta A_{FeI} / \Delta \chi_{exc}$ = 0.000 $\pm$ 0.002 dex/eV
(excitation equilibrium) and $A_{FeI} = A_{FeII}$
(ionization equilibrium).

The results are shown in Figure 1, where the difference
between physical and spectroscopic parameters ($\Delta$ \teff, $\Delta$
log $g$)
are plotted as a function of \teff and log $g$.
As can be seen, we succeed to bring into rough agreement
the temperatures determined from the IRFM and excitation
equilibrium of FeI lines. However, ionization equilibrium
is not satisfied, and spectroscopic gravities could be
off by as much as 0.4 dex, which is much higher than the
uncertainties expected from {\it Hipparcos} parallaxes
($<$ 0.01 dex) or errors in masses and bolometric corrections
(a few 0.01 dex). 

In the lower panel we show a plot of $\Delta$ log $g$ vs. $\Delta$ \teff.
If spectroscopic equilibrium is fulfilled in a LTE 1D analysis,
($\Delta$ log $g$, $\Delta$ \teff) are expected to be distributed 
randomly inside an ellipse with semi-axis about 50 K 
in \teff and 0.05 dex in log $g$, centered in (0, 0) 
(provided the physical atmospheric parameters are 
correct in an absolute scale). More conservatively, we considered
that if spectroscopic parameters are correct they should
satisfy: ($\Delta$ \teff /75 K)$^2$ + ($\Delta$ log $g$/0.075 dex)$^2$ $<$ 1.
However, even considering these very generous errors, there are
no points inside the expected
region (Fig. 1, lower panel), and a strong correlation between the errors in
spectroscopic \teff and log $g$ is observed,
showing that spectroscopic atmospheric parameters are not independent. 
Note that this can not be
solved by invoking large errors in the absolute scale of $gf$-values
of FeII lines, since those errors would affect the
zero point of $\Delta$ log $g$, but not both 
the unacceptable  large range in error (about 0.4 dex) 
and the correlation between $\Delta$ log $g$ and $\Delta$ \teff.

\begin{figure}
\plotone{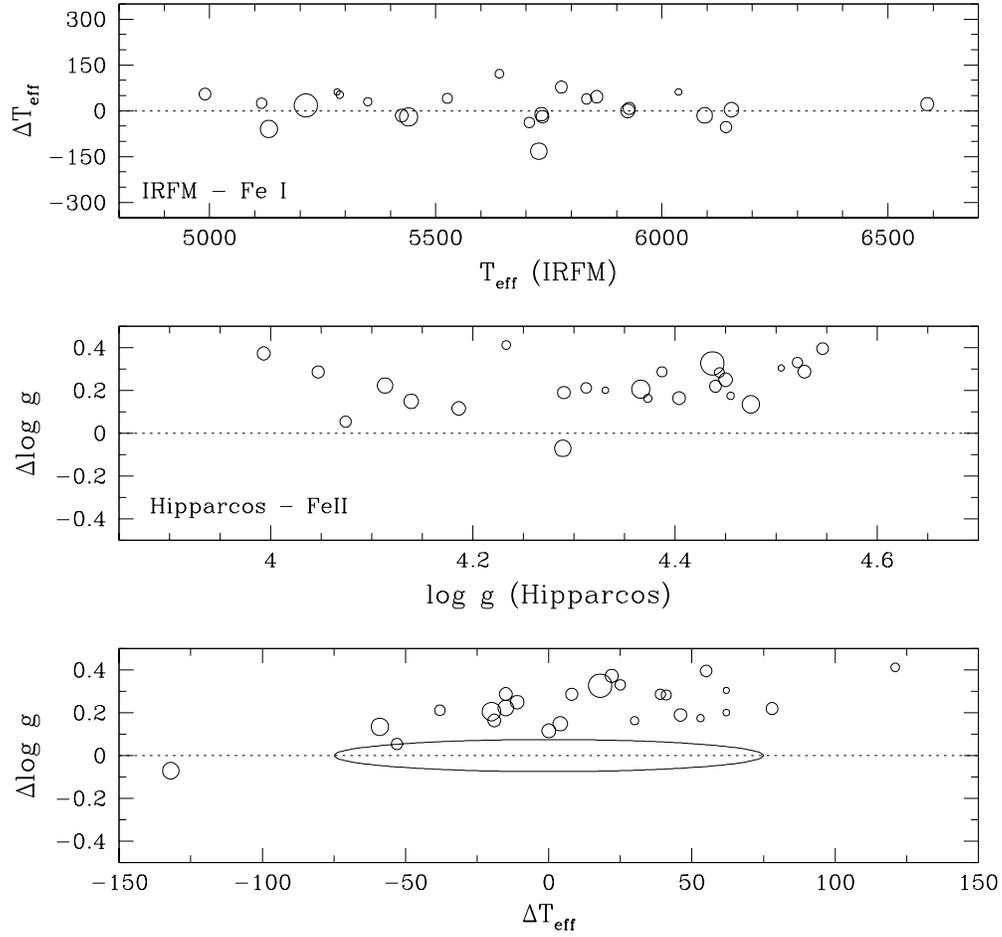}
\caption{Differences between the physical (IRFM + {\it Hipparcos})
and spectroscopic atmospheric parameters derived in this work.
The size of the circles is proportional to their metallicities.
The zone inside the ellipse (lower panel),
($\Delta$ \teff /75 K)$^2$ + ($\Delta$ log $g$/0.075 dex)$^2$ $<$ 1,
shows the expected region of consistency 
between physical and spectroscopic parameters.}
\label{tg}
\end{figure}

\section{Conclusions}
Classical LTE 1D modeling may be failing 
to satisfy spectroscopic equilibrium
of iron lines, and the atmospheric parameters estimated from 
this assumption may be wrong. Since FeII is the main
ionization stage in G and F dwarfs and it is
expected to be formed in LTE, the problem is probably
due to NLTE effects on FeI lines.
Considering NLTE on line formation of FeI lines
results in a higher iron abundance than
that derived from a LTE approach (e.g. Shchukina \& Trujillo Bueno 2001), 
thus lowering the discrepancy observed between spectroscopic and
{\it Hipparcos} gravities. 

A detailed discussion, as well as 
implications for the metallicity distribution
of stars with planets will be presented 
in a forthcoming article (Mel\'endez \& Ram\'{\i}rez 2005, in preparation).

\acknowledgements
JM acknowledges support from NFS grant AST-0205951 to JGC.
This publication have made use of data from the Hipparcos astrometric mission 
of the ESA.

\end{document}